\documentclass[aps,prd,a4paper,preprintnumbers,amsmath,amssymb,superscriptaddress,twocolumn,floatfix,showpacs]{revtex4}
\usepackage{amssymb}
\usepackage{float,graphicx}

\newcommand{\Pbar}{\bar{\mathcal{P}}_{\gamma \leftrightarrow \phi}}
\newcommand{\be}{\begin{eqnarray}}
\newcommand{\ee}{\end{eqnarray}}
\newcommand{\ba}{\left( \begin{array}{ccc}}
\newcommand{\ea} {\end{array} \right)}
\newcommand{\bv}{\left( \begin{array}{c}}
\newcommand{\ev} {\end{array} \right)}

\begin{document}
\title{Chameleon-Photon Mixing in a Primordial Magnetic Field}
\author{Camilla A.O. Schelpe}
\email{C.A.O.Schelpe@damtp.cam.ac.uk}
\affiliation{Department of Applied Mathematics and Theoretical Physics,
Centre for Mathematical Sciences,  Cambridge CB3 0WA, United Kingdom}
\date{\today}
\begin{abstract}
The existence of a sizable, $\mathcal{O}\left(10^{-10}\text{--}10^{-9}\mathrm{G}\right)$, cosmological magnetic field in the early Universe has been postulated as a necessary step in certain formation scenarios for the large scale $\mathcal{O}(\mu\mathrm{G})$ magnetic fields found in galaxies and galaxy clusters. If this field exists then it may induce significant mixing between photons and axion-like particles (ALPs) in the early Universe. The resonant conversion of photons into ALPs in a primordial magnetic field has been studied elsewhere by Mirizzi, Redondo and Sigl (2009). Here we consider the non-resonant mixing between photons and scalar ALPs with masses much less than the plasma frequency along the path, with specific reference to the chameleon scalar field model. The mixing would alter the intensity and polarization state of the cosmic microwave background (CMB) radiation. We find that the average modification to the CMB polarization modes is negligible. However the average modification to the CMB intensity spectrum is more significant and we compare this to high precision measurements of the CMB monopole made by the far infrared absolute spectrophotometer (FIRAS) on board the COBE satellite. The resulting 95\% confidence limit on the scalar-photon conversion probability in the primordial field (at $100\,\mathrm{GHz}$) is $\Pbar<2.6\times10^{-2}$. This corresponds to a degenerate constraint on the photon-scalar coupling strength, $g_{\mathrm{eff}}$, and the magnitude of the primordial magnetic field. Taking the upper bound on the strength of the primordial magnetic field derived from the CMB power spectra, $B_{\lambda}\leq 5.0\times 10^{-9}\mathrm{G}$, this would imply an upper bound on the photon-scalar coupling strength in the range  $g_{\mathrm{eff}}\lesssim 7.14\times10^{-13}\mathrm{GeV}^{-1}$ to $g_{\mathrm{eff}}\lesssim 9.20\times10^{-14}\mathrm{GeV}^{-1}$, depending on the power spectrum of the primordial magnetic field. 
\end{abstract}
\maketitle

\section{Introduction\label{intro}}

The origin of the observed large-scale magnetic fields of order $\mu\mathrm{G}$  found in nearly all galaxies and galaxy clusters is still largely unknown. It is generally believed that the galactic magnetic field develops by some form of amplification from a pre-galactic cosmological magnetic field. The two popular formation scenarios that are considered are either some exponential dynamo mechanism which amplifies a very small seed field of order $10^{-30}\mathrm{G}$ as the galaxy evolves, or the adiabatic collapse of a larger existing cosmological field of order $(10^{-10}\text{--}10^{-9})\mathrm{G}$. There are a number of pros and cons to both scenarios. See \cite{Kronberg94} for reviews on the subject. Theories explaining the origin of this primordial magnetic (PMF) field are still highly speculative. It has been suggested that a large-scale magnetic field could be produced during inflation if the conformal invariance of the electromagnetic field is broken; see for example \cite{Ratra92} or more recently \cite{Kunze10}. 

To date, there is no astrophysical evidence for the existence of a large-scale cosmological magnetic field, and only upper bounds on its magnitude have been derived. So far the strongest constraints have come from measurements of the cosmic microwave background (CMB) and big-bang nucleosynthesis  \cite{Kronberg94,Tashiro06,Kosowsky09,Yamazaki10,Paoletti10}. The CMB bounds are derived by considering the effects of Faraday rotation induced by the PMF on the CMB power spectra. In \cite {Kosowsky09} an upper limit in the range $6\times10^{-8}$ to $2\times10^{-6}\mathrm{G}$ was derived for the mean-field amplitude of the PMF at a comoving length scale of $1\,\mathrm{Mpc}$ by comparison to the WMAP 5-year data. More recent work \cite{Yamazaki10,Paoletti10} analysing the WMAP data in combination with other CMB experiments such as ACBAR, CBI and QUAD have placed tighter constraints on the amplitude of the PMF with an upper bound on the mean-field amplitude at $1\,\mathrm{Mpc}$ of $\sim 5\times10^{-9}\mathrm{G}$.

The existence of a primordial magnetic field would induce mixing between CMB photons and axion-like particles (ALPs). ALPs refer collectively to any very light scalar or pseudo-scalar with a linear coupling to $F_{\mu\nu}F^{\mu\nu}$ or $\epsilon_{\mu\nu\rho\sigma}F^{\mu\nu}F^{\rho\sigma}$ respectively. In this paper we consider non-resonant mixing of scalar ALPs and CMB photons, with specific reference to the chameleon scalar field model \cite{Khoury04,Brax07,Burrage08}. Standard ALPs have constant mass and photon-scalar coupling everywhere, while the chameleon model has a density dependent mass. In sparse environments, such as the primordial plasma, the chameleon acts as a very light scalar field and would be indistinguishable from a standard scalar ALP. However in dense environments the chameleon is very heavy and evades the standard ALP constraints \cite{Brax09}. The current best constraints on the coupling strength between the chameleon and electromagnetic fields are: $g_{\mathrm{eff}} \lesssim 9.1\times 10^{-10} \mathrm{GeV}^{-1}$ \cite{Burrage08} and $g_{\mathrm{eff}} \lesssim \left(0.72\sim22\right)\times 10^{-9} \mathrm{GeV}^{-1}$ \cite{Davis09}. Other chameleon-like theories exist such as the Olive-Pospelov model \cite{Davis09,Olive08} which have a density-dependent coupling strength, but we do not discuss these further here.  

The resonant mixing of photons with scalar and pseudo-scalar ALPs in a primordial magnetic field has been analysed by Mirizzi, Redondo and Sigl \cite{Mirizzi09}. They find a constraint on the combined magnetic field strength, $B$, and ALP-photon coupling strength, $g$: $g\langle B^2 \rangle ^{1/2}\lesssim 10^{-13}\sim10^{-11}\mathrm{GeV}^{-1}\mathrm{nG}$, for ALP masses between $10^{-14}\mathrm{eV}$ and $10^{-4}\mathrm{eV}$. We believe that these results will not necessarily be applicable to the chameleon because its mass evolves as the density of the Universe decreases. 

In the following analysis we assume a stochastic primordial magnetic field with a power-law power spectrum, similar to the treatment in \cite{Kosowsky09}. We assume fluctuations in the magnetic field are damped on small scales due to Alfv$\mathrm{\acute{e}}$n wave dissipation \cite{Barrow98}, and subdivide the magnetic field into multiple domains of length $L$ of a comparable size to just above the Alfv$\mathrm{\acute{e}}$n wave damping scale $k_{D}^{-1}$. The magnetic field in each domain is assumed to be approximately constant, and correlated to the other domains according to the magnetic power spectrum. A similar method was applied to correlations in quasar polarization spectra by Agarwal, Kamal and Jain \cite{Agarwal09}.    

The degree of conversion between chameleons and photons in a magnetic field is inversely proportional to the electron density in the plasma \cite{Davis09}. Hence the dominant contribution to photon-scalar mixing will take place in the region after recombination when the ionization fraction drops significantly and before reionization. This greatly simplifies the mixing equations because we do not need to evolve the photon-scalar mixing equations through the last scattering surface, nor include the density inhomogeneities present after reionization. We model a scenario in which the primary CMB is formed at the last scattering surface and then evolves through a primordial magnetic field extending from recombination ($z\sim1100$) to the epoch of reionization ($z\sim20$). 

This paper is organized as follows: in section \ref{chameleon_model} the chameleon model is introduced and the calculations describing photon-scalar mixing in a magnetic field, living in a Friedman-Robertson-Walker (FRW) spacetime, are presented. The power spectrum of the primordial magnetic field is discussed in more detail in section \ref{magnetic_field}. In section \ref{evolution} we analyse the evolution of the photon and chameleon states as they propagate through the multiple magnetic domains, and predict the average modification to the CMB intensity and polarization. In section \ref{experiment}, our predictions are compared to precision measurements of the CMB monopole made by the far infrared absolute spectrophotometer (FIRAS) on board the cosmic background explorer (COBE) satellite. We present a summary of the work and our conclusions in section \ref{conclusion}. The appendix contains details of the equations governing the evolution of the CMB Stokes parameters.

\section{Chameleon-Photon Mixing in an Expanding Universe\label{chameleon_model}}

The chameleon scalar field has been suggested as a candidate for the
dark energy quintessence field \cite{Khoury04}. Standard quintessence scalar fields
have coupling strengths to matter which are unnaturally fine-tuned
to very low values so as to be compatible with fifth-force experiments.
By contrast the chameleon model is constructed so that the chameleon
can have a gravitational strength (or greater) coupling to normal matter
while at the same time evading fifth-force constraints. This is achieved
by introducing a density-dependent term in the effective potential
of the chameleon, which causes a change to the minimum of the potential
depending on the density of the surrounding matter. In sparse environments
the chameleon behaves as an effectively massless scalar field while
in a laboratory on Earth it is much heavier and evades detection. 

In addition to the matter coupling, the chameleon can have a non-zero coupling, $\frac{\phi}{M_{\mathrm{eff}}}F_{\mu\nu}F^{\mu\nu}$, to the electromagnetic (EM) field. The mass parameter, $M_{\mathrm{eff}}$, describing the strength of the photon coupling is best constrained by observations of radiation passing through astrophysical magnetic fields. The two best current constraints in terms of the coupling strength $g_{\rm eff}=1/M_{\rm eff}$ were presented in section \ref{intro}. In terms of $M_{\rm eff}$, the lower limit from considering constraints on the production of starlight polarization in the galactic magnetic field is $M_{\mathrm{eff}} \gtrsim 1.1\times 10^9 \mathrm{GeV}$ \cite{Burrage08}. Measurements of the Sunyaev--Zel'dovich effect in galaxy clusters places a lower bound on $M_{\mathrm{eff}}$ in the range $4.5\times 10^7 \mathrm{GeV}$ to $1.4\times 10^9 \mathrm{GeV}$, depending on the model assumed for the cluster magnetic field \cite{Davis09}.

The action describing the chameleon model is that of a generalized
scalar-tensor theory:\begin{eqnarray*}
\mathcal{S} & = & \int\mathrm{d}^{4}x\sqrt{-g}\left(\frac{1}{2}M_{Pl}^{2}\mathcal{R}-\frac{1}{2}g^{\mu\nu}\partial_{\mu}\phi\partial_{\nu}\phi\right.\\
& & \left.-V(\phi)-\frac{1}{4}B_{F}(\phi/M)F_{\mu\nu}F^{\mu\nu}\right)\\
 &  & +\mathcal{S}_{\mathrm{matter}}\left(\psi^{(i)},\, g_{\mu\nu}^{(i)}\right)\,,\end{eqnarray*}
where $V(\phi)$ is the self-interaction potential of the scalar field $\phi$, and $\mathcal{S}_{\mathrm{matter}}$ is the matter action, excluding the
kinetic term of electromagnetism, which contains the coupling of the
matter fields $\psi^{(i)}$ to the conformal metric $g_{\mu\nu}^{(i)}\equiv B_{i}^{2}(\phi/M)g_{\mu\nu}$.
The $B_{i}(\phi/M)$ determine the coupling of the scalar field to
different matter species, and $B_{F}(\phi/M)$ determines the photon-scalar
coupling. For simplicity a universal matter coupling, $B_{i}(\phi/M)=B_{\mathrm{m}}(\phi/M)$, is assumed.
We take the Universe to be described by a spatially-flat Friedman-Robertson-Walker
(FRW) metric: $\mathrm{d}s^{2}=-\mathrm{d}t^{2}+a^{2}(t)\left(\mathrm{d}x^{2}+\mathrm{d}y^{2}+\mathrm{d}z^{2}\right)$ 
where $a(t)$ is the time-dependent
scale factor describing the expansion, normalized to $a_{0}=1$ today. The coordinates $\left(x,\,y,\,z\right)$ are comoving, related to the physical coordinates by $\tilde{x}_i=ax_i$, and $t$ is the proper time. 

The equation of motion for the $\phi$ field is found by varying $\mathcal{S}$
with respect to $\delta\phi$: 
\begin{eqnarray*}
\square\phi & \equiv & -\partial_{t}^{2}\phi-3H(t)\partial_{t}\phi+\frac{1}{a^{2}}\nabla^{2}\phi \\
& =& V^{\prime}(\phi)+\frac{1}{4}F^{2}B_{F}^{\prime}(\phi)-B_{i}^{3}(\phi)B_{i}^{\prime}(\phi)T_{\mathrm{m}}^{(i)},
\end{eqnarray*}
where $H(t)\equiv\dot{a}/a$ is the Hubble expansion rate. $T_{\mathrm{m}}^{(i)}$
is the trace of the stress-energy tensor in the conformal frame described
by the metric $g_{\mu\nu}^{(i)}$, for which particle masses are constant
and independent of $\phi$. This is related to the stress-energy tensor
in the physical frame by $T_{\mathrm{m}}=B_{i}^{3}T_{\mathrm{m}}^{(i)}$.
In general, $T_{\mathrm{m}}=-(\rho_{\mathrm{m}}-3P_{\mathrm{m}})$
corresponding to the physical, measured density and pressure. If we
assume the chameleon does not couple to the energy density of dark matter or any other exotic particles, then $T_{\mathrm{m}}=-\rho_{\mathrm{b}}$,
the baryonic matter density. Hence \[
\square\phi=V_{\mathrm{eff}}^{\prime}(\phi;\, F^{2},\,\rho_{b}),\]
where, \begin{eqnarray*}
V_{\mathrm{eff}}(\phi;\, F^{2},\,\rho_{b}) & \equiv & V(\phi)+\frac{1}{4}F^{2}B_{F}(\phi)+\rho_{b}B_{\mathrm{m}}(\phi).\end{eqnarray*}
The form of the self-interaction potential $V(\phi)$ determines whether
a general scalar-tensor theory is chameleon-like or not. For a chameleon field we require the potential
to be of runaway form. A typical choice of potential can be described by 
\begin{equation}
V(\phi)\approx\Lambda_{0}+\frac{\Lambda^{n+4}}{n\phi^{n}}\,,\label{eq:self-interaction potential}\end{equation}
where for reasons of naturalness $\Lambda\sim\mathcal{O}\left(\Lambda_{0}\right)$.
If the chameleon is to be a suitable candidate for dark energy, we require
$\Lambda_{0}=\left(2.4\pm0.3\right)\times10^{-3}\mathrm{eV}$. The
chameleon mass is defined by 
\[ m_{\phi}^{2}\equiv V_{,\phi\phi}^{\mathrm{eff}}\left(\phi_{\mathrm{min}};\,\bar{F}^{2},\,\rho_{b}\right)\,,\]
where $\bar{F}^{\mu\nu}$ is the background value of the electromagnetic
field tensor. 

The photon equation of motion comes from varying $\mathcal{S}$
with respect to $\delta A_{\mu}$:\[
\nabla_{\nu}\left(B_{\mathrm{F}}(\phi)F^{\mu\nu}\right)=0\,.\]
Any electromagnetic components in $L_{\mathrm{matter}}$ would
introduce electromagnetic currents on the right-hand side of the equation. We neglect these at this stage in the calculation. The primordial plasma is a good conductor and so any currents will be small \cite{Barrow07}. To a good approximation the induced currents arising from photon propagation will be described by the plasma frequency which is included later in this calculation. 

In standard chameleon theories we assume $\phi/M\ll1$ and approximate
$B_{\mathrm{F}}\approx1+\phi/M_{\mathrm{eff}}$ and $B_{\mathrm{m}}^{2}\approx1+2\phi/M$,
where this defines $M_{\mathrm{eff}}$ and $M$. We expect these two
coupling strengths to be of similar magnitude but do not require it.
Under this approximation, and splitting the electromagnetic field into
the background EM field $\bar{F}^{\mu\nu}$ and the photon field $f^{\mu\nu}$,
the above equations of motion become 
\begin{eqnarray*}
\square\varphi & \simeq & m_{\phi}^{2}\varphi+\frac{1}{4M_{\mathrm{eff}}}\left(f_{\mu\nu}\bar{F}^{\mu\nu}+f^{\mu\nu}\bar{F}_{\mu\nu}\right)\,,\\
\nabla_{\mu}f^{\mu\nu} & \simeq & -\frac{1}{M_{\mathrm{eff}}}\left(\nabla_{\mu}\varphi\right)\bar{F}^{\mu\nu}\,,\end{eqnarray*}
where $\varphi\equiv\phi-\bar{\phi}$ is the perturbation in the scalar
field about its background value, and we neglect terms that are $\mathcal{O}\left(f^{2}\right)$ and
$\mathcal{O}(\phi f^{\mu\nu}/M_{\rm eff})$.

In an inertial, locally Minkowskian frame, with metric $\hat{g}_{\mu\nu}=\mathrm{diag}\left(-1,1,1,1\right)$,
the electromagnetic field tensor has the form \[
\hat{F}^{\mu\nu}=\left(\begin{array}{cccc}
0 & E_{x} & E_{y} & E_{z}\\
-E_{x} & 0 & B_{z} & -B_{y}\\
-E_{y} & -B_{z} & 0 & B_{x}\\
-E_{z} & B_{y} & -B_{x} & 0\end{array}\right)\,.\]
The coordinate transformation from the locally Minkowski metric, $\hat{g}_{\mu\nu}=\mathrm{diag}\left(-1,1,1,1\right)$,
into the FRW metric, $g_{\mu\nu}=\mathrm{diag}\left(-1,\, a^{2},\, a^{2},\, a^{2}\right)$, is \[
F^{\mu\nu}=\Lambda_{\,\alpha}^{\mu}\Lambda_{\,\beta}^{\nu}\hat{F}^{\alpha\beta}\,,\]
with 
\[
\Lambda_{\,\nu}^{\mu}=\frac{\partial x^{\mu}}{\partial\hat{x}^{\nu}}=\mathrm{diag}\left(1,1/a,1/a,1/a\right).\]
Thus in an FRW expanding Universe the EM field tensor is given by
\[
F^{\mu\nu}=\left(\begin{array}{cccc}
0 & E_{x}/a & E_{y}/a & E_{z}/a\\
-E_{x}/a & 0 & B_{z}/a^{2} & -B_{y}/a^{2}\\
-E_{y}/a & -B_{z}/a^{2} & 0 & B_{x}/a^{2}\\
-E_{z}/a & B_{y}/a^{2} & -B_{x}/a^{2} & 0\end{array}\right)\,.\]
We take the background to be a large-scale
magnetic field $\mathbf{B}$, and assume the contributions
from electric fields in the plasma are negligible. The photon field
can be described by the quantum polarization states $a^{\mu}$ where
$f_{\mu\nu}\equiv\nabla_{\mu}a_{\nu}-\nabla_{\nu}a_{\mu}$. Taking
the Lorentz gauge condition $\nabla_{\mu}a^{\mu}=\partial_{\mu}a^{\mu}+3Ha^{0}=0$,
we find the equations of motion become 
\begin{eqnarray*}
-\partial_{t}^{2}\varphi-3H(t)\partial_{t}\varphi+\frac{1}{a^{2}}\nabla^{2}\varphi & = & m_{\phi}^{2}\varphi+\frac{\mathbf{B}\cdot(\boldsymbol{\nabla}\times\mathbf{a})}{M_{\mathrm{eff}}}\,,
\end{eqnarray*}
and
\begin{eqnarray*}
-\partial_{t}^{2}\mathbf{a}-5H\partial_{t}\mathbf{a}+2qH^{2}\mathbf{a}-4H^{2}\mathbf{a}+\frac{1}{a^{2}}\nabla^{2}\mathbf{a} & = & \frac{\boldsymbol{\nabla}\varphi\times\mathbf{B}}{a^{2}M_{\mathrm{eff}}},\end{eqnarray*}
where we define the deceleration parameter, $q(t)\equiv-\ddot{a}a/\dot{a}^{2}$.

In this analysis we have assumed that the background values of the
chameleon and photon fields are slowly varying over length and time
scales of $\mathcal{O}(1/\omega)$ where $\omega$ is the proper frequency
of the electromagnetic radiation being considered. We further assume
that the frequency of the radiation satisfies $\omega\gg H$ and that
$q\lesssim\mathcal{O}(1)$. The equations describing mixing between
the photons and scalar field are then \begin{eqnarray*}
-\ddot{\mathbf{a}}+\frac{1}{a^{2}}\nabla^{2}\mathbf{a} & \simeq & \frac{\boldsymbol{\nabla}\varphi\times\mathbf{B}}{a^{2}M_{{\rm eff}}}+\omega_{{\rm pl}}^{2}\mathbf{a}\,,\\
-\ddot{\varphi}+\frac{1}{a^{2}}\nabla^{2}\varphi & \simeq & \frac{\mathbf{B}\cdot(\boldsymbol{\nabla}\times\mathbf{a})}{M_{{\rm eff}}}+m_{\phi}^{2}\varphi\,,\end{eqnarray*}
where we have included the plasma frequency, $\omega_{{\rm pl}}^{2}=4\pi\alpha_{{\mathrm{em}}}n_{\rm e}/m_{\rm e}$, as an effective photon mass. Whenever electromagnetic radiation propagates through a plasma with electron number density
$n_{\mathrm{e}}$ it displaces the electrons slightly and induces oscillations at their natural frequency $\omega_{\rm pl}$. This interaction of the EM wave with the electron density hinders its progress and acts as an effective mass for the photons. 

Taking the radiation to be propagating in the $\mathbf{\hat{z}}$
direction of an orthonormal Cartesian basis $(\mathbf{\hat{x}},\mathbf{\hat{y}},\mathbf{\hat{z}})$,
with $\mathbf{a}=(\gamma_{x},\gamma_{y},0)^{{\rm T}}$, the equations
of motion for the chameleon and photon can be written in matrix form:
\[
\left[-\partial_{t}^{2}+\frac{\partial_{z}^{2}}{a^{2}}-\left(\begin{array}{ccc}
\omega_{{\rm pl}}^{2} & 0 & -\frac{B_{y}\partial_{z}}{a^{2}M_{{\rm eff}}}\\
0 & \omega_{{\rm pl}}^{2} & \frac{B_{x}\partial_{z}}{a^{2}M_{{\rm eff}}}\\
\frac{B_{y}\partial_{z}}{M_{{\rm eff}}} & -\frac{B_{x}\partial_{z}}{M_{{\rm eff}}} & m_{\phi}^{2}\end{array}\right)\right]\left(\begin{array}{c}
\vert\gamma_{x}\rangle\\
\vert\gamma_{y}\rangle\\
\vert\varphi\rangle\end{array}\right)=0\,.\]
Note how the chameleon scalar field only mixes with the component of photon polarization aligned
perpendicular to the transverse magnetic field. The magnitude of the magnetic field aligned parallel to the photon path ($B_z$) plays no part in the mixing.

Following a similar procedure to that in \cite{Davis09,Raffelt88}
we assume the fields vary slowly over time compared to the frequency
of the radiation and that the refractive index is close to unity,
which requires $m_{{\rm \phi}}^{2}/2\omega^{2}$, $\omega_{{\rm pl}}^{2}/2\omega^{2}$
and $\vert B\vert/2\omega M_{{\rm eff}}$ all $\ll1$. Defining $\vert\gamma_{i}\rangle=\vert\hat{\gamma}_{i}(z)\rangle e^{i\omega(az-t)}$
and $\vert\varphi\rangle=\vert\hat{\varphi}(z)\rangle e^{i\omega(az-t)-i\beta}$, we approximate $-\partial_{t}^{2}\approx\omega^{2}$ and $\partial_z\approx i\omega a$ such that
$\omega^{2}+\frac{1}{a^{2}}\partial_{z}^{2}\approx2\omega(\omega+i\frac{1}{a}\partial_{z})$. Thus,  
\[
\left[i\frac{\omega}{a}\partial_{z}-\left(\begin{array}{ccc}
\omega_{{\rm pl}}^{2} & 0 & -\frac{iB_{y}\omega}{2aM_{{\rm eff}}}\\
0 & \omega_{{\rm pl}}^{2} & \frac{iB_{x}\omega}{2aM_{{\rm eff}}}\\
\frac{iB_{y}\omega a}{2M_{{\rm eff}}} & -\frac{iB_{x}\omega a}{2M_{{\rm eff}}} & m_{\phi}^{2}\end{array}\right)\right]\left(\begin{array}{c}
\vert\hat{\gamma}_{x}\rangle\\
\vert\hat{\gamma}_{y}\rangle\\
\vert\hat{\varphi}\rangle\end{array}\right)=0\,.\]
We further simplify the mixing matrix by defining $\vert\hat{\gamma}_{i}\rangle=\vert\bar{\gamma_{i}}(z)\rangle e^{-i\beta}$
and $\vert\hat{\varphi}\rangle=\vert\bar{\varphi}(z)\rangle e^{-i\beta}$ with $\partial_{z}\beta\equiv a\omega_{\rm pl}^2/2\omega$. The magnetic field and frequency can be expressed in terms of their comoving values: $\mathbf{B}_{0}=a^{2}\mathbf{B}$
and $\omega_{0}=a\omega$. In the subsequent analysis we drop the subscript-zero notation for comoving magnetic
field values. Thus we can write 
\begin{eqnarray}
\left[ia\partial_{z}-\left(\begin{array}{ccc}
0 & 0 & \frac{-B_{y}}{2M_{{\rm eff}}}\\
0 & 0 & \frac{B_{x}}{2M_{{\rm eff}}}\\
\frac{-B_{y}}{2M_{{\rm eff}}} & \frac{B_{x}}{2M_{{\rm eff}}} & \frac{a^{3}m_{\mathrm{eff}}^{2}}{2\omega_{0}}\scriptstyle{-i\partial_{z}a}\end{array}\right)\right]\left(\begin{array}{c}
\vert\bar{\gamma}_{x}\rangle\\
\vert\bar{\gamma}_{y}\rangle\\
\frac{1}{a}\vert i\bar{\varphi}\rangle\end{array}\right)=0, \label{eq:mixing matrix}\end{eqnarray}
where $m_{\mathrm{eff}}^{2}\equiv m_{\phi}^{2}-\omega_{\mathrm{pl}}^{2}$.

To solve this system of equations we must make various simplifying assumptions. We neglect spatial fluctuations in the electron density along the path length, and only consider a simple redshift scaling: $n_{\rm e}(z)\simeq n_{\rm b0}X_e a^{-3}$, where $X_e$ is the ionization fraction along the path. The ionization fraction drops rapidly at recombination from  $X_e\simeq1$ to its final freeze-out value $X_e\sim 5\times 10^{-4}$ \cite{Seager99}, and only increases again at the epoch of reionization. The conversion between photons and light scalar particles in a magnetic plasma scales inversely with the electron number density in the plasma \cite{Davis09}. Thus the conversion will be suppressed for high values of the ionization fraction. We assume the dominant contribution to photon-scalar mixing in a primordial magnetic field occurs between redshift $\sim750$ (by which time the ionization fraction has already dropped to $10^{-3}$)  and redshift $\sim20$ (onset of reionization). We approximate the ionization fraction as being  at a constant value of $X_e\sim5\times10^{-4}$ in this region, and neglect contributions from other sections of the path length.       

The chameleon mass for the generalized self-interaction potential of Eq. (\ref{eq:self-interaction potential}) is \begin{eqnarray*}
m_{\phi}^{2} & \simeq & (n+1)\Lambda_{0}^{-\frac{n+4}{n+1}}\left(\frac{\rho_{\mathrm{b}0}}{a^{3}M}+\frac{\vert\mathbf{B}\vert^{2}}{2a^{4}M_{\mathrm{eff}}}\right)^{\frac{n+2}{n+1}}\\
 & \approx & 8.4\times10^{-57}a^{-9/2}\left(\frac{M_{\mathrm{eff}}}{10^{9}\mathrm{GeV}}\right)^{-\frac{3}{2}}\mathrm{GeV}^{2},\end{eqnarray*}
where the second line assumes $n=\mathcal{O}(1)$, a magnetic field of less than $10^{-9}\mathrm{G}$, a chameleon coupling strength of less than $10^{-9}\mathrm{GeV}^{-1}$, and that the average baryonic density is 4\% of the critical density. We compare this to the plasma frequency, \begin{eqnarray*}
\omega_{\mathrm{pl}}^{2} & = & \frac{4\pi\alpha_{\mathrm{em}}n_{\mathrm{e}}}{m_{\mathrm{e}}}=\left(\frac{4\pi\alpha_{\mathrm{em}}X_{\mathrm{e}}\rho_{b0}}{m_{\mathrm{e}}m_{b}}\right)a^{-3} \equiv  p_{0}a^{-3}\\
 & \approx & 1.73\times10^{-49}a^{-3}\,\mathrm{GeV}^{2}\end{eqnarray*}
over the redshift range $\sim 20$ to 750, where we have taken the average mass per baryon to be $m_{b}\simeq937\,\mathrm{MeV}$ \cite{Steigman06}. It is clear then that $\omega_{\rm pl}^2\gg m_{\phi}^2$ along the path from recombination to reionization. Our subsequent analysis requires the chameleon to be sufficiently light for the plasma frequency to dominate over the path length, and thus is not dependent on the specific form of the chameleon potential. It applies to any scalar ALP satisfying this condition. 

The comoving distance, $z$, travelled by the photons is related to the value of the scale factor at $z$ by \[
\mathrm{d}z = \mathrm{d}a/a^{2}H(a)\,.\] This leads to \[
a(z) \simeq \left(a^{1/2}(0)+\frac{1}{2}H_{0}\Omega_{m0}^{1/2}z\right)^{2}\] in the matter dominated era between recombination and reionization,
where $\Omega_{m0}= 0.26$ \cite{Dunkley08}. Over the total path length from a redshift of $750$ to $20$, the comoving distance covered is $L_{\mathrm{tot}}\simeq2.1\, h^{-1}\mathrm{Gpc}$.

Changing the integration variable in Eq. (\ref{eq:mixing matrix}) to \[
\xi(z)\equiv\frac{-2}{H_{0}\Omega_{m0}^{1/2}}a^{-\frac{1}{2}}(z)\,,\]
such that $\partial_{\xi}=a\partial_{z}$, and noting $p_0/2\omega_0\gg\partial_za$,
we find \begin{eqnarray}
\left[i\partial_{\xi}-
\left(\begin{array}{ccc}
0 & 0 & \frac{-B_{y}(\xi)}{2M_{{\rm eff}}}\\
0 & 0 & \frac{B_{x}(\xi)}{2M_{{\rm eff}}}\\
\frac{-B_{y}(\xi)}{2M_{{\rm eff}}} & \frac{B_{x}(\xi)}{2M_{{\rm eff}}} & -\frac{p_{0}}{2\omega_{0}}\end{array}\right)
\right]\left(\begin{array}{c}
\vert\bar{\gamma}_{x}\rangle\\ \vert\bar{\gamma}_{y}\rangle\\ \vert\bar{\chi}\rangle\end{array}\right) = 0
\label{eq:mixing eq}\end{eqnarray}
where 
$\vert\bar{\chi}\rangle\equiv\frac{1}{a}\vert i\bar{\varphi}\rangle$. The $\xi$--dependence of the mixing matrix is entirely due to fluctuations in the magnetic field.

\section{The Primordial Magnetic Field\label{magnetic_field}}

The generic model for the primordial magnetic field is of a stochastic field parameterized by a power-law power spectrum up to a cut-off scale $k_D$, \[
P(k)=A_{B}k^{n_{B}},\;\;\;\;k<k_D,\] where $A_B$ is some normalization constant. The spectral index $n_B$ must be greater than $-3$ to prevent infrared divergences in the integral over the power spectrum at long wavelengths. Assuming a statistically homogeneous and isotropic field, the power spectrum is defined by
\[
\langle B_{i}(\mathbf{k})B_{j}^{\star}(\mathbf{k}')\rangle\equiv(2\pi)^{3}P_{ij}P(k)\delta(\mathbf{k}-\mathbf{k}')\,,\]
where $P_{ij}\equiv\delta_{ij}-\hat{k}_{i}\hat{k}_{j}$ is the projector
onto the transverse plane imposed by the divergence-free nature of
the magnetic field. We adopt the Fourier transform convention, \[
B_{i}(\mathbf{k})=\int\mathrm{d}^{3}\mathbf{x}\,B_{i}(\mathbf{x})e^{-i\mathbf{k}\cdot\mathbf{x}}.\]The two-point correlation function is then 
\[
\langle B_{i}(\mathbf{x})B_{j}(\mathbf{y})\rangle=\frac{1}{(2\pi)^{3}}\int\mathrm{d}^{3}\mathbf{k}P_{ij}P(k)e^{i\mathbf{k}\cdot(\mathbf{x}-\mathbf{y})}.\]
Following the prescription in \cite{Kosowsky09,Kosowsky05}, normalization of the magnetic field is achieved by convolving the field (in real space) with a Gaussian smoothing kernel of comoving
radius $\lambda_{B}$. Defining the Fourier transform of the Gaussian
smoothing kernel to be $f(\mathbf{k})=e^{-\lambda_{B}^{2}k^{2}/2}$, the Fourier transform of the convolved field is \[
B_{i}(\mathbf{k})\vert_{\lambda_{B}}=B_{i}(\mathbf{k})\cdot f(\mathbf{k})\,.\]
The mean-field amplitude of the smoothed field, $B_{\lambda}$, is given by \begin{eqnarray*}
B_{\lambda}^{2} & = & \langle B_{i}(\mathbf{x})\vert_{\lambda_{B}}B_{i}(\mathbf{x})\vert_{\lambda_{B}}\rangle\\
 & = & \frac{2A_{B}}{(2\pi)^{2}}\frac{1}{\lambda^{n_{B}+3}}\Gamma\left(\frac{n_{B}+3}{2}\right).\end{eqnarray*}
Hence the normalization constant, \[
A_{B}=\frac{(2\pi)^{n_{B}+5}B_{\lambda}^{2}}{2k_{\lambda}^{n_{B}+3}\Gamma\left(\frac{n_{B}+3}{2}\right)}\,,\]
where $k_{\lambda}=2\pi/\lambda_{B}$. 

Constraints on the magnitude of the primordial magnetic field from a comparison of Faraday rotation effects in the CMB with the WMAP 5-year data were given in \cite{Kosowsky09}. They found that the upper limit on the mean-field amplitude of the magnetic field on a comoving length scale of $\lambda_{B}=1\,\mathrm{Mpc}$ was in the range $6\times10^{-8}$ to $2\times10^{-6}\mathrm{G}$ ($95\%\:\mathrm{CL}$) for a spectral index $n_B=-2.9$ to $-1$. This range for the spectral index was based on the likely formation scenarios for the primordial magnetic field \cite{Ratra92} and current exclusion bounds on the spectral index \cite{Caprini01}. More recent results analysing the WMAP 5-year data in combination with other CMB experiments \cite{Yamazaki10} constrained the mean-field amplitude $B_{1\,\mathrm{Mpc}}<2.98\times 10^{-9}\mathrm{G}$ and the spectral index $n_B<-0.25$ ($95\%\:\mathrm{CL}$). An analysis of the latest WMAP 7-year data \cite{Paoletti10} derived upper bounds of $B_{1\,\mathrm{Mpc}}<5.0\times 10^{-9}\mathrm{G}$ and $n_B<-0.12$ ($95\%\:\mathrm{CL}$).

The magnetic field power spectrum will be damped at small length-scales. We assume a cut-off to the power spectrum at the Alfv$\mathrm{\acute{e}}$n wave damping scale \cite{Barrow98}. This damping scale is approximated \cite{Kosowsky05} as \[
\left(\frac{k_{D}}{\mathrm{Mpc}^{-1}}\right)^{n_{B}+5}\approx2.9\times10^{4}h\left(\frac{B_{\lambda}}{10^{-9}\mathrm{G}}\right)^{-2}\left(\frac{k_{\lambda}}{\mathrm{Mpc}^{-1}}\right)^{n_{B}+3}\]
where $H_{0}\equiv100h\,\mathrm{km\, s^{-1}Mpc^{-1}}$.

Using these definitions, we can calculate the correlations in the magnetic field along the line of sight: 
\begin{eqnarray*}
\langle
B_{i}(x\hat{\mathbf{z}})B_{j}(y\hat{\mathbf{z}})\rangle & =
&\frac{1}{(2\pi)^3}\intop_{0}^{2\pi}\intop_{0}^{\pi}\intop_{0}^{\infty}k^2\sin\theta_k\mathrm{d}k\mathrm{d}\theta_k\mathrm{d}\phi_k\\
& & \cdot\left(\delta_{ij}-\frac{k_ik_j}{k^2}\right) P(k) e^{i(x-y)k\cos\theta_k},\end{eqnarray*}
where we have written the wave vector in spherical polar coordinates, $\mathbf{k} = \left(k,\theta_k,\phi_k\right)$. Thus 
\begin{eqnarray*}
\langle B_{x}(x\hat{\mathbf{z}})B_{y}(y\hat{\mathbf{z}})\rangle = \langle B_{y}(x\hat{\mathbf{z}})B_{x}(y\hat{\mathbf{z}})\rangle = 0, \end{eqnarray*}
and 
\begin{eqnarray*}
\langle B_{x}(x\hat{\mathbf{z}})B_{x}(y\hat{\mathbf{z}})\rangle=\langle B_{y}(x\hat{\mathbf{z}})B_{y}(y\hat{\mathbf{z}})\rangle\equiv R_{\mathrm{B}}(x-y),\end{eqnarray*}
where \begin{eqnarray*}
R_{\mathrm{B}}(z) = \frac{2}{(2\pi)^2}\intop_{0}^{\infty}k^{2}P(k)\mathrm{d}k \left[\frac{\sin(zk)}{zk}+\frac{\cos(zk)}{(zk)^2}-\frac{\sin(zk)}{(zk)^3}\right].
\end{eqnarray*}
Substituting the assumed form for the magnetic power spectrum, and changing variables, we find
\begin{eqnarray}
R_{\mathrm{B}}(x-y) = \frac{B_{\lambda}^{2}K\left[(x-y)k_D\right]}{\Gamma\left(\frac{n_B+3}{2}\right)}\left(\frac{2\pi k_{D}}{k_{\lambda}}\right)^{n_{B}+3},\label{RB}\end{eqnarray}
with \begin{eqnarray} 
K\left[\theta\right]\equiv \intop_{0}^{1}t^{n_{B}+2}\mathrm{d}t\left[\frac{\sin(\theta t)}{\theta t}+\frac{\cos(\theta t)}{(\theta t)^{2}}-\frac{\sin(\theta t)}{(\theta t)^{3}}\right]. \label{integralK}
\end{eqnarray}

\section{Evolution of the Stokes Parameters\label{evolution}}

Following a similar procedure to that in \cite{Agarwal09} we subdivide the path length into many small domains of length $L$,
each with a constant magnetic field strength and direction, and take correlations in the field across the different domains to be described by the magnetic power spectrum. The inverse power-law form assumed for the primordial magnetic power spectrum imposes a rapidly decreasing amplitude to the field fluctuations and an even faster decreasing slope to the magnetic power spectrum with increasing $k$. At very large $k$ the amplitude of the fluctuations are damped to zero by Alfv${\rm \acute{e}}$n wave dissipation. However slightly above the Alfv${\rm \acute{e}}$n wave damping scale, $k_{D}^{-1}$, the amplitude of the field is non-zero yet the slope of the power spectrum will be approximately flat. In what follows we assume on scales slightly greater than $k_{D}^{-1}$ the magnetic field can be approximated as constant and take $L\sim \mathcal{O}\left(k_{D}^{-1}\right)$. The exact value of $L$ is only necessary in determining the number of magnetic domains that the photons traverse along the path. Some variation in $L$ will not affect the order of magnitude of the final results. 

\subsection{Evolution through a Single Domain\label{single_evolution}}

For a single domain we define $B$ to be the magnitude of the transverse component of the magnetic field, and $\sigma$ to be the angle it makes with
the $x$-axis. Within the domain these stay approximately constant. Returning to Eq. (\ref{eq:mixing eq}), we rotate the $\left(\bar{\gamma}_{x},\bar{\gamma}_{y}\right)$
basis so as to reduce the problem to two-component mixing.  Defining  \begin{eqnarray*}
\vert\bar{\gamma}_{a}\rangle & = & \cos\sigma\vert\bar{\gamma}_{x}\rangle+\sin\sigma\vert\bar{\gamma}_{y}\rangle\,,\\
\vert\bar{\gamma}_{b}\rangle & = & -\sin\sigma\vert\bar{\gamma}_{x}\rangle+\cos\sigma\vert\bar{\gamma}_{y}\rangle, \end{eqnarray*} we find $\vert\bar{\gamma}_{a}\rangle$ is constant over the domain and \[
\left[i\partial_{\xi}-\left(\begin{array}{cc}
0 & B/2M\\
B/2M & -p_{0}/2\omega_{0}\end{array}\right)\right]\left(\begin{array}{c}
\vert\bar{\gamma}_{b}\rangle\\
\vert\bar{\chi}\rangle\end{array}\right)=0.\]
Diagonalisation of this two-component mixing leads to \begin{eqnarray*}
\vert\bar{\gamma}_{b}^{\rm new}(L)\rangle & = & \vert\bar{\gamma}_{b}^{\rm new}(0)\rangle e^{-i\left(\Delta-\Delta/\cos2\theta\right)}\,,\\
\vert\bar{\chi}^{\rm new}(L)\rangle & = & \vert\bar{\chi}^{\rm new}(0)\rangle e^{-i\left(\Delta+\Delta/\cos2\theta\right)}\,,\end{eqnarray*}
where\begin{eqnarray*}
\vert\bar{\gamma}_{b}^{\rm new}\rangle & = & \cos\theta\vert\bar{\gamma}_{b}\rangle+\sin\theta\vert\bar{\chi}\rangle\,,\\
\vert\bar{\chi}^{\rm new}\rangle & = & -\sin\theta\vert\bar{\gamma}_{b}\rangle+\cos\theta\vert\bar{\chi}\rangle\,,\end{eqnarray*}
and \begin{eqnarray}
\tan2\theta & \equiv & \frac{2\omega_{0}B}{p_{0}M_{\mathrm{eff}}},\label{theta}\\
\Delta & \equiv & -\frac{p_{0}}{4\omega_{0}}\left(\xi(L)-\xi(0)\right) \approx -\frac{p_{0}L}{4a\omega_{0}}\,. \label{Delta}\end{eqnarray}
Additionally defining \begin{eqnarray*}
A & \equiv & \sin2\theta\sin\left(\frac{\Delta}{\cos2\theta}\right)\,,\\
\tan\psi & \equiv & \cos2\theta\tan\left(\frac{\Delta}{\cos2\theta}\right),\end{eqnarray*}
the chameleon and photon polarization
states in the rotated basis after passing through a single domain of length $L$, are \begin{eqnarray*}
\vert\bar{\gamma}_{a}(L)\rangle & = & \vert\bar{\gamma}_{a}(0)\rangle\,,\\
\vert\bar{\gamma}_{b}(L)\rangle & = & e^{i\alpha}\sqrt{1-A^{2}}\vert\bar{\gamma}_{b}(0)\rangle+ie^{-i\Delta}A\vert\bar{\chi}(0)\rangle\,,\\
\vert\bar{\chi}(L)\rangle & = & e^{-i\beta}\sqrt{1-A^{2}}\vert\bar{\chi}(0)\rangle+ie^{-i\Delta}A\vert\bar{\gamma}_{b}(0)\rangle\,,\end{eqnarray*}
where $\alpha=\psi-\Delta$ and $\beta=\psi+\Delta$.
The probability of conversion between chameleons and photons in a single domain is $P=A^{2}$.

The intensity and polarization state of radiation is best described
by its Stokes parameters: intensity, $I$, linear polarization, $Q$ and $U$, and circular polarization, $V$. These are defined in terms of the photon polarization states by \begin{eqnarray*}
I & = & \langle\gamma_{x}\vert\gamma_{x}\rangle+\langle\gamma_{y}\vert\gamma_{y}\rangle\,, \\
Q & = & \langle\gamma_{x}\vert\gamma_{x}\rangle-\langle\gamma_{y}\vert\gamma_{y}\rangle\,, \\
U +iV& = & 2\langle\gamma_{x}\vert\gamma_{y}\rangle\,. \end{eqnarray*}
We additionally define chameleon `Stokes parameters' to close the evolution equations, 
\begin{eqnarray*}
J+iK & = & 2e^{i\psi}\langle\gamma_{x}\vert\chi\rangle\,, \\
L+iM & = & 2e^{i\psi}\langle\gamma_{y}\vert\chi\rangle\,, \\
I_{\chi} & = & \langle\chi\vert\chi\rangle\,.\end{eqnarray*}

The evolution of the Stokes parameters through a single magnetic domain,
in the rotated $\left(\bar{\gamma}_{a},\,\bar{\gamma}_{b}\right)$ basis, is then \begin{eqnarray*}
\tilde{I}_{\gamma} & \rightarrow & \left(1-\frac{A^{2}}{2}\right)\tilde{I}_{\gamma}+\frac{A^{2}}{2}\tilde{Q}+A^{2}\tilde{I}_{\chi} \\ 
& & +A\sqrt{1-A^{2}}\left(\sin2\psi \tilde{L}-\cos2\psi \tilde{M}\right)\,,\\
\tilde{Q} & \rightarrow & \left(1-\frac{A^{2}}{2}\right)\tilde{Q}+\frac{A^{2}}{2}\left(\tilde{I}_{\gamma}-2\tilde{I}_{\chi}\right)\\
& & -A\sqrt{1-A^{2}}\left(\sin2\psi \tilde{L}-\cos2\psi \tilde{M}\right)\,,\\
\tilde{U}+i\tilde{V} & \rightarrow & e^{i\alpha}\sqrt{1-A^{2}}\left(\tilde{U}+i\tilde{V}\right)\\
& & +ie^{-i\beta}A\left(\tilde{J}+i\tilde{K}\right)\,,\end{eqnarray*}
and \begin{eqnarray*}
\tilde{J}+i\tilde{K} & \rightarrow & e^{-i\beta}\sqrt{1-A^{2}}\left(\tilde{J}+i\tilde{K}\right)\\
& & +ie^{i\alpha}A\left(\tilde{U}+i\tilde{V}\right)\,, \\
\tilde{L}+i\tilde{M} & \rightarrow & e^{-2i\psi}\left(1-A^{2}\right)\left(\tilde{L}+i\tilde{M}\right)\\
& & +e^{2i\psi}A^{2}\left(\tilde{L}-i\tilde{M}\right)\\
& & +iA\sqrt{1-A^{2}}\left(\tilde{I}_{\gamma}-\tilde{Q}-2\tilde{I}_{\chi}\right)\,.\end{eqnarray*}
$I_{\chi}$ is found by requiring that the total flux of photons and chameleons along the path length is conserved, $I_{\chi}=\left(I_{\mathrm{tot}}-I_{\gamma}\right)/a^2$.

\subsection{Evolution through Multiple Magnetic Domains\label{multiple_evolution}}

The evolution equations above can be extended to many domains following a similar procedure to that
in \cite{Burrage08}. 

For CMB photons in the range $30\text{--}600\,\mathrm{GHz}$, passing
through a magnetic field of less than $10^{-9}\mathrm{G}$, and assuming
a photon-scalar coupling strength no greater than $10^{-9}\mathrm{GeV^{-1}}$, the mixing parameters defined in Eqs. (\ref{theta}) and (\ref{Delta}) satisfy $\theta\ll1$
and $\vert\Delta\vert\gg1$ in a domain of length $L\sim k_D^{-1}$. There is a region towards the end of the path at larger $a$ for which $\Delta \sim 1$, but the approximations that follow involving averaging terms in $\sin\Delta$ and $\cos\Delta$ along the path are still 
valid in this instance. Thus on average over the total path length, \[A^{2}\simeq2\theta^{2}\ll1\,.\] This places us in the regime of weak-mixing which can be solved analytically if we neglect terms smaller than $\mathcal{O}(NA^2)$, where $N=L_{\mathrm{tot}}/L$ is the number of magnetic domains along the path. Details of the mixing equations in multiple domains are presented in the Appendix. Here we quote the average, over many lines of sight, of the modification to the Stokes parameters. We assume there is no initial flux of chameleons and define $z_{n}\simeq nL$ to be the location of the $n^{\mathrm{th}}$ domain. Then,  
\begin{eqnarray*}
\frac{\langle\delta I_{\gamma}\rangle}{\langle I_{0}\rangle} & \simeq & -\left(\frac{2\omega_{0}}{p_{0}M_{\mathrm{eff}}}\right)^{2}\left(\frac{1}{2}NR_{\mathrm{B}}(0)\right.\\
& & \left.+\sum_{n=0}^{N-1}\sum_{r=0}^{n-1}\cos2\Delta(n-r)R_{\mathrm{B}}(z_{r}-z_{n})\right)\,,
\end{eqnarray*}
\begin{eqnarray*}
\frac{\langle\delta Q\rangle}{\langle Q_0\rangle} & \simeq & \frac{\langle\delta U\rangle}{\langle U_0\rangle}\\ 
&\simeq & -\left(\frac{2\omega_{0}}{p_{0}M_{\mathrm{eff}}}\right)^{2}\left(\frac{1}{2}NR_{\mathrm{B}}(0)\right.\\
& & \left.+\sum_{n=0}^{N-1}\sum_{r=0}^{n-1}\cos2\Delta(n-r)R_{\mathrm{B}}(z_{r}-z_{n})\right)\\
& & -\Delta^{2}\left(\frac{2\omega_{0}}{p_{0}M_{\mathrm{eff}}}\right)^{4}\left(\frac{1}{2}N\left[R_{\mathrm{B}}(0)\right]^{2}\right.\\
& & \left.+\sum_{n=0}^{N-1}\sum_{r=0}^{n-1}\left[R_{\mathrm{B}}(z_{r}-z_{n})\right]^{2}\right)\,,
\end{eqnarray*}
\begin{eqnarray*}
\frac{\langle\delta V\rangle}{\langle V_0\rangle} & \simeq & -\left(\frac{2\omega_{0}}{p_{0}M_{\mathrm{eff}}}\right)^{2}\left(\frac{1}{2}NR_{\mathrm{B}}(0)\right.\\
& & \left.+\sum_{n=0}^{N-1}\sum_{r=0}^{n-1}\cos2\Delta(n-r)R_{\mathrm{B}}(z_{r}-z_{n})\right) \\
& & -2\Delta^{2}\left(\frac{2\omega_{0}}{p_{0}M_{\mathrm{eff}}}\right)^{4}N\left[R_{\mathrm{B}}(0)\right]^{2},\end{eqnarray*}
where $R_{\mathrm{B}}(x-y)$ was defined in Eq (\ref{RB}). Over the range $n_B=-2.9$ to $-1$, the integral $K(\theta)$ (Eq. (\ref{integralK})) can be well approximated by a window function: $W(\theta)=2/3(n_{B}+3)$ for $\vert\theta\vert\lesssim2$, and zero otherwise. 
Taking the domain length $L\sim k_{D}^{-1}$, the average modification to the CMB intensity is then 
\begin{eqnarray}
\frac{\left\langle \delta I_{\gamma}\right\rangle}{\left\langle I_{0}\right\rangle}  & \approx & \frac{-NB_{\lambda}^{2}\left(2\pi k_{D}/k_{\lambda}\right)^{n_{B}+3} }{3(n_B+3)\Gamma\left(\frac{n_B+3}{2}\right)}\left(\frac{2\omega_{0}}{p_{0}M_{\mathrm{eff}}}\right)^{2}\,,\label{eq:Intensity} \end{eqnarray}
where we have averaged terms in $\cos\Delta$ and $\sin\Delta$, since $\left|\Delta\right|\gg1$. Of the average modifications to the Stokes parameters, that of the CMB intensity will be most significant since the CMB is only weakly polarized with $\langle V_0^2\rangle^{1/2}\ll\langle Q_0^2\rangle^{1/2},\langle U_0^2\rangle^{1/2}\ll I_0$.

In addition to the average modification to the Stokes parameters as a result of photon-scalar mixing, there will also be a change to the correlations between different Stokes parameters. A full derivation of the cross-correlations along different lines of sight would require us to expand the primordial field power spectrum in spherical harmonics, a task we reserve for a future paper \cite{Schelpe10}. However we can estimate the magnitude of the effect by looking at, say, the $\langle UV\rangle$ cross-correlation along the line of sight. The dominant contribution, neglecting terms proportional to $\langle Q_{0}U_{0}\rangle$, $\langle I_{0}V_{0}\rangle$ and smaller, is \begin{eqnarray*}
& & \langle UV\rangle \approx -\textstyle\left(\frac{2\omega_{0}}{p_{0}M_{\mathrm{eff}}}\right)^{4} \cdot\left(\frac{1}{2}\Delta\displaystyle{\sum_{n,m=0}^{N-1}}\scriptstyle\left[R_{\mathrm{B}}\left(z_{n}-z_{m}\right)\right]^{2}\right.\\
& & +\sum_{n,m=0}^{N-1}\sum_{r=0}^{n-1}\sum_{s=0}^{m-1}\scriptstyle\cos2\Delta(n-r)\sin2\Delta(m-s)R_{\mathrm{B}}\left(z_{n}-z_{m}\right)R_{\mathrm{B}}\left(z_{r}-z_{s}\right)\\
& & \left.+\Delta\sum_{n,m=0}^{N-1}\sum_{r=0}^{n-1}\scriptstyle R_{\mathrm{B}}\left(z_{n}-z_{m}\right)R_{\mathrm{B}}\left(z_{r}-z_{m}\right)\cos2\Delta(n-r)\right)\langle I_{0}Q_{0}\rangle\\
& & \approx -2N\Delta\textstyle\left(\frac{B_{\lambda}^{2}\left(2\pi k_{D}/k_{\lambda}\right)^{n_{B}+3} }{(n_B+3)\Gamma\left(\frac{n_B+3}{2}\right)}\right)^2\left(\frac{2\omega_{0}}{p_{0}M_{\mathrm{eff}}}\right)^{4} \langle I_{0}Q_{0}\rangle.\end{eqnarray*}
From this we see that a significant $\langle UV\rangle$ correlation can arise from the much larger $\langle IQ\rangle$ correlation. Although the above analysis only applies to the average correlations along many lines of sight, for which $\langle I_{0}Q_{0}\rangle\simeq0$ in the CMB, we expect a similar magnitude to the effect when a full derivation for the cross-correlations along different lines of sight is made. It opens up the exciting possibility of a significant chameleon contribution to the $\langle EB\rangle$ power spectrum of the CMB.

\section{Bounds on the CMB Intensity from FIRAS\label{experiment}}

Returning to Eq. (\ref{eq:Intensity}), we have that the average modification to the CMB intensity from scalar-photon mixing
in a primordial magnetic field is
\begin{eqnarray*}
\frac{\left\langle \delta I_{\gamma}\right\rangle }{I_{\gamma0}}\equiv-\bar{\mathcal{P}}_{\gamma\rightarrow\phi} & \approx & -\left(\frac{\omega_{0}B_{\mathrm{eff}}}{p_{0}M_{\mathrm{eff}}}\right)^{2},\end{eqnarray*}
where we have defined
\begin{eqnarray*}
B_{\mathrm{eff}}^{2} & \equiv & \frac{4}{3}N\frac{B_{\lambda}^{2}\left(\lambda_{B}k_{D}\right)^{n_{B}+3}}{\left(n_{B}+3\right)\Gamma\left(\frac{n_{B}+3}{2}\right)}\,.\end{eqnarray*}
This expresses the primordial magnetic field in terms of its mean-field amplitude, $B_{\lambda}$, at a length-scale of $\lambda_B$. Most constraints on the magnitude of the magnetic field are expressed in terms of these parameters. The value of $n_B$ dictates the slope of the magnetic power spectrum, and we consider the range $-2.9$ to $-1$. 

\begin{figure}
\includegraphics[clip,width=7.5cm]{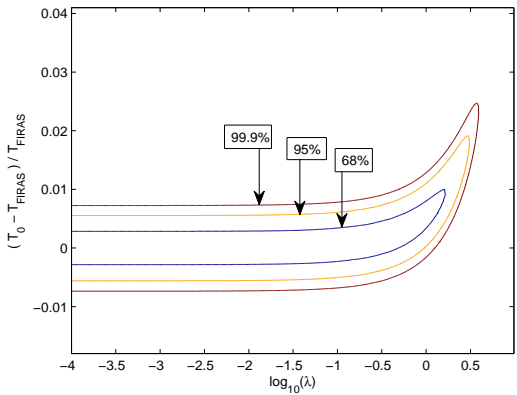}\caption{\label{Confidence-limits}Confidence limits on the black-body
temperature, $T_0$, and the strength of photon-scalar mixing, parameterized by $\lambda$, coming from precision measurements of the CMB monopole, where $T_{\mathrm{FIRAS}}=2.725\,\mathrm{K}$.}

\end{figure}

The most precise measurements to date of the CMB monopole near the peak in its spectrum come from the far infrared
absolute spectrophotometer (FIRAS) on board the cosmic background explorer
(COBE) satellite \cite{Fixsen96,Fixsen02}. These measurements fit exceptionally
closely to the spectrum of a black-body at a temperature of $2.725\mathrm{K}\pm1\mathrm{mK}$. In Table 4 of \cite{Fixsen96}, the residuals of the CMB monopole that remain after
subtracting foreground effects are listed. These residuals, added to the spectrum
of a perfect black-body radiating at $2.725\,\mathrm{K}$, can be
compared to the predictions of photon-scalar mixing to constrain the
parameter space. The FIRAS instrument has 43 significant
frequency channels in the range $60\text{--}630\,\mathrm{GHz}$ after
calibration. The spectral intensity of a black-body at temperature
$T_{0}$ is
\[
I_{0}\left(\nu,T_{0}\right)=\frac{4\pi\nu^{3}}{\exp\left(2\pi\nu/T_{0}-1\right)}\,.\]
In the presence of photon-scalar mixing the spectrum is modified to
\begin{equation}
I\left(\nu,T_{0},\lambda\right)=\left(1-\bar{\mathcal{P}}_{\gamma\rightarrow\phi}\right)I_{0}\left(\nu,T_{0}\right)\,,\label{eq:pred-ch}\end{equation}
where $\lambda$ is defined such that \begin{eqnarray*}
\bar{\mathcal{P}}_{\gamma\rightarrow\phi} & \simeq & 2.7\lambda\times10^{-2}\left(\frac{\nu}{100\,\mathrm{GHz}}\right)^{2},\\
\lambda\left(B_{\mathrm{eff}},\, M_{\mathrm{eff}}\right) & = & \left(\frac{B_{\mathrm{eff}}}{10^{-9}\mathrm{G}}\right)^{2}\left(\frac{M_{\mathrm{eff}}}{10^{9}\mathrm{GeV}}\right)^{-2}.\end{eqnarray*}
We maximize the likelihood, $L$, defined by \[
-2\log L=\sum_{i=1}^{43}\left(\frac{I^{obs}(\nu_{i})-I(\nu_{i},T_{0},\lambda)}{\sigma^{obs}(\nu_{i})}\right)^{2},\] over the parameter space $\left(T_{0},\lambda\right)$. Here $I^{obs}$ is the observed CMB monopole measured by FIRAS at 43 different frequencies, $\nu_{i}$, with
standard errors $\sigma^{obs}$. $I(\nu_{i},T_{0},\lambda)$ is the spectral intensity predicted
for chameleon-photon mixing given in Eq (\ref{eq:pred-ch}). We treat $T_{0}$ as a free parameter in our analysis rather
than fixing it at $T_{\mathrm{FIRAS}}=2.725\,\mathrm{K}$.
Confidence limits are estimated by assuming \[
\chi^{2}=-\log\left(\frac{L\left(T_{0},\lambda\right)}{L\left(\hat{T}_{0},\hat{\lambda}\right)}\right)\]
follows a $\chi_{1}^{2}$ distribution, where $\left(\hat{T}_{0},\hat{\lambda}\right)$
are the best fit values of $T_{0}$ and $\lambda$ in the parameter
space, found by the maximum likelihood procedure. Minimizing $\chi^{2}$
with respect to $\lambda$ gives $T_{0}=2.725\pm0.016\,\mathrm{K}$,
identical to the value obtained when no chameleon mixing is considered.
Minimizing with respect to $T_{0}$ gives the following 95\% confidence
limit on $\lambda$,
\[
\log_{10}\lambda<-0.01\;\;(95\%).\]
Fig.\;\ref{Confidence-limits} shows the 68\%, 95\% and 99.9\%
confidence limits on the $\left(T_{0},\lambda\right)$ parameter space.

\begin{figure}\includegraphics[clip,width=7.5cm]{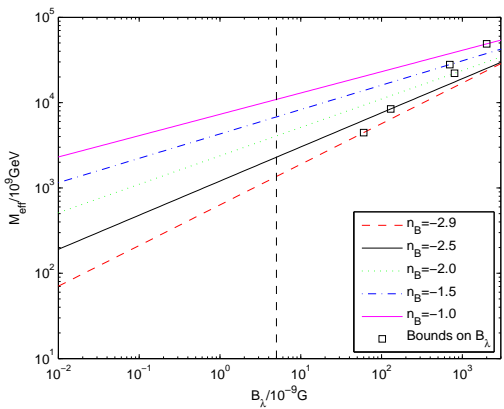}\caption{\label{fig:MBbounds} Exclusion bounds on the primordial magnetic field and photon-scalar coupling strength, $g_{\mathrm{eff}}=1/M_{\mathrm{eff}}$, from precision measurements of the CMB monopole, at the 95\% confidence level. $B_{\lambda}$ is the mean-field amplitude of the magnetic field at a comoving length-scale of $1\,\mathrm{Mpc}$. Open squares correspond to the upper limit on the primordial magnetic field found in \cite{Kosowsky09}, while the vertical line corresponds to the upper limit found in \cite{Yamazaki10,Paoletti10}. The region to the bottom right of the plot is excluded for different values of the magnetic spectral index $n_B$.}

\end{figure}

The limit on $\lambda$ corresponds to a degenerate constraint on the magnitude of the primordial magnetic field and the chameleon-photon coupling strength. The magnetic field strength is expressed in terms of the mean-field amplitude at a comoving length-scale of $1\,\mathrm{Mpc}$, $B_{\lambda}$. In Fig.\;\ref{fig:MBbounds} we plot the exclusion bounds in the $\left(B_{\lambda},M_{\mathrm{eff}}\right)$ parameter space resulting from the 95\% confidence limit on $\lambda$. Different lines correspond to different values of the magnetic spectral index, $n_B$, in the range -2.9 to -1. The open squares mark the 95\% upper limit on the allowed values for $B_{\lambda}$ (for each $n_B$) found in \cite{Kosowsky09} from comparison of Faraday rotation effects in the CMB with WMAP 5-year data. The vertical line corresponds to the more recent constraints on $B_{\lambda}$ given in \cite{Yamazaki10} and \cite{Paoletti10}. The region to the bottom right of the plot for each $n_B$ value is excluded at the 95\% confidence level. For example if we take a magnetic field strength at the upper limit allowed by Faraday rotation effects in the CMB, $B_{\lambda}\simeq5\times10^{-9}\mathrm{G}$, the corresponding bound on the photon-scalar coupling strength, $g_{\mathrm{eff}}=1/M_{\mathrm{eff}}$, is
\[M_{\mathrm{eff}}\gtrsim \left(0.14\text{--}1.09\right)\times 10^{13}\mathrm{GeV}\,,\] 
depending on the slope of the magnetic power spectrum. 

\section{Conclusions\label{conclusion}}

The existence of a large-scale cosmological field in the early Universe is so far unconfirmed. It would need to be in the region of $\mathcal{O}\left(10^{-10}\sim10^{-9}\mathrm{G}\right)$ if it is to explain the formation of the $\mathcal{O}(\mu\mathrm{G})$ magnetic fields in galaxies and galaxy clusters, through adiabatic collapse. If this primordial magnetic field exists it would induce mixing between CMB photons and axion-like particles (ALPs) as they propagate from the last-scattering surface to Earth. 

In this paper we have studied the case of non-resonant mixing between scalar-ALPs and photons in a primordial magnetic field, with specific reference to the chameleon scalar field model. 

The chameleon model is a promising candidate for the dark energy scalar field since it can have a gravitational strength (or stronger) coupling to normal matter while at the same time evading fifth-force constraints. To date, the strongest bounds on the chameleon coupling strength come from chameleon-photon mixing in local astrophysical environments, such as starlight propagating through the galactic magnetic field: $g_{\mathrm{eff}}\lesssim 9.1\times 10^{-10}$ \cite{Burrage08}. Should there be a detection of a primordial magnetic field of order $\gtrsim 10^{-10}\mathrm{G}$, our results would place far greater constraints on the coupling strength.         

We have considered a stochastic primordial magnetic field described by a power-law power spectrum, $P(k)\propto k^{n_B}$, up to some cut-off damping scale $k_D$. Photon-scalar mixing in this field was solved by dividing the path length into multiple magnetic domains in which the field is approximated as constant. The length of the domains was taken to be of a comparable size to $k_D ^{-1}$ since the magnetic power spectrum will be approximately flat, but non-zero, just above the damping scale. Correlations between the magnetic field strength and direction in each domain are determined by the magnetic power spectrum. In addition, we approximated the ionization fraction as being constant in the region from redshift $\sim750$ to $20$, and held at its post-recombination freeze-out value of $X_e\sim 5\times 10^{-4}$. The dominant contribution to photon-scalar mixing in the CMB will occur in this region of low electron density, and we neglected contributions from other elements of the path length. A more sophisticated approach to modelling the electron density from recombination to the present day may result in small changes to the predictions, but would require a numerical rather than analytical approach to the mixing equations. 

We have compared our predictions of the average modification of the CMB intensity over the whole sky, to precision measurements of the CMB monopole by the FIRAS instrument on board the COBE satellite \cite{Fixsen96,Fixsen02}. This constrains the probability of photon-scalar mixing over the path length, to be \[\bar{\mathcal{P}}_{\gamma\rightarrow\phi}\lesssim 0.026\;\;(95\%\:\mathrm{CL})\] at $100\,\mathrm{GHz}$. The corresponding bounds on the magnitude of the magnetic field and photon-scalar coupling strength are plotted in Fig. \ref{fig:MBbounds} for different values of the magnetic spectral index. Until a detection of the primordial magnetic field is made, to break the degeneneracy of this constraint, we cannot place limits on the chameleon-photon coupling strength, $g_{\rm eff}$. For the largest magnetic field allowed by the constraints in \cite{Yamazaki10,Paoletti10} we would find the strongest possible constraint, \[ g_{\mathrm{eff}}\lesssim \left(0.92 \sim 7.14\right)\times10^{-13}\mathrm{GeV}^{-1},\] depending on the slope of the magnetic field power spectrum.  

The results in this paper apply to any scalar ALP with a mass less than $\sim 10^{-14}\mathrm{eV}$, since we require the mass to be much smaller than the plasma frequency along the path. These nicely complement the bounds derived in \cite{Mirizzi09} for resonant conversion between photons and ALPs in a primordial magnetic field, $g_{\rm eff}\langle B^2 \rangle^{1/2}\lesssim 10^{-13}\sim 10^{-11}\,{\rm GeV^{-1}\,nG}$, which apply to ALP masses in the range $10^{-14}{\rm eV}$ to $10^{-4}{\rm eV}$.   

In addition to the average modification to the CMB intensity, the formalism developed in this paper can straightforwardly be extended to calculate the change to correlations between the CMB Stokes parameters along different lines of sight. In section \ref{multiple_evolution}, an example was given of how a correlation between the $U$ and $V$ polarization modes can arise from photon-scalar mixing in the primordial magnetic field, given a non-zero $\langle IQ\rangle$ correlation. If this effect is present in the CMB cross-correlations, it would lead to a significant chameleon signature in the $\langle EB\rangle$ power spectrum. A full analysis of this effect is kept for a separate publication \cite{Schelpe10}.

\vspace{0.5cm}

\noindent{\bf Acknowledgments:} I am funded by STFC. I am grateful to Douglas Shaw and my supervisor Anne Davis for their support, and to  Anthony Challinor for interesting and helpful discussions.

\appendix
\section{Photon-Scalar Mixing in Multiple Magnetic Domains}
In section \ref{single_evolution} we found the evolution equations of the Stokes parameters after passing through a single magnetic domain of length $L$, within which the magnetic field strength and direction are approximated as constant. This can be extended to $N$ multiple domains, under the assumption of weak-mixing for which we require $NA^2\ll1$ and $N\gg1$. CMB photons in the range $30\text{--}600\,\mathrm{GHz}$, propagating through a weak primordial magnetic field, satisfy $\theta\ll1$ and
$\vert\Delta\vert\gg1$. In this limit $\psi\simeq\Delta$, $\beta\simeq2\Delta$ and $\alpha\sim\mathcal{O}(A)\ll1$.
We define $\delta I_{n+1}$ to be
the difference in intensity after passing through the $n^{th}$ domain
compared to its initial value, $I_{0}$,
and similarly for the other parameters. Neglecting terms smaller than
$\mathcal{O}(A^{2})$, we obtain the following recurrence relations
for the Stokes parameters:
\begin{eqnarray*}
\left(\delta I_{\gamma}\right)_{n+1} & \simeq & \left(\delta I_{\gamma}\right)_{n}-\frac{A_{n}^{2}}{2}I_{0}-\frac{A_{n}^{2}}{2}\left(\cos2\sigma_{n}Q_{0}+\sin2\sigma_{n}U_{0}\right)\\
 &  & +A_{n}\cos\sigma_{n}\left[L_{n}\sin2\Delta-M_{n}\cos2\Delta\right]\\
 &  & +A_{n}\sin\sigma_{n}\left[K_{n}\cos2\Delta-J_{n}\sin2\Delta\right]\,,\\
\delta Q_{n+1} & \simeq & \delta Q_{n}-\frac{A_{n}^{2}}{2}Q_{0}+\frac{A_{n}^{2}}{2}\cos2\sigma_{n}I_{0}+\alpha_{n}\sin2\sigma_{n}V_{n}\\
& & -\frac{\alpha_{n}^{2}}{2}\sin2\sigma_{n}\left(\sin2\sigma_{n}Q_{0}-\cos2\sigma_{n}U_{0}\right)\\
 &  & -A_{n}\cos\sigma_{n}\left[L_{n}\sin2\Delta-M_{n}\cos2\Delta\right]\\
 &  & +A_{n}\sin\sigma_{n}\left[K_{n}\cos2\Delta-J_{n}\sin2\Delta\right]\,,\\
\delta U_{n+1} & \simeq  & \delta U_{n}-\frac{A_{n}^{2}}{2}U_{0}+\frac{A_{n}^{2}}{2}\sin2\sigma_{n}I_{0}-\alpha\cos2\sigma_{n}V_{n}\\
& & -\frac{\alpha_{n}^{2}}{2}\cos2\sigma_{n}\left(\cos2\sigma_{n}U_{0}-\sin2\sigma_{n}Q_{0}\right)\\
 &  & -A_{n}\sin\sigma_{n}\left[L_{n}\sin2\Delta-M_{n}\cos2\Delta\right]\\
 &  & -A_{n}\cos\sigma_{n}\left[K_{n}\cos2\Delta-J_{n}\sin2\Delta\right]\,,\\
\delta V_{n+1} & \simeq & \delta V_{n}-\frac{A_{n}^{2}}{2}V_{0}-\frac{\alpha_{n}^{2}}{2}V_{0}\\ 
& & +\alpha_{n}\left(\cos2\sigma_{n}U_{n}-\sin2\sigma_{n}Q_{n}\right)\\
 &  & +A_{n}\sin\sigma_{n}\left[L_{n}\cos2\Delta+M_{n}\sin2\Delta\right]\\
 &  & +A_{n}\cos\sigma_{n}\left[K_{n}\sin2\Delta+J_{n}\cos2\Delta\right]\,,\end{eqnarray*}
with \begin{eqnarray*}
J_{n+1}+iK_{n+1} & \simeq & e^{-2i\Delta}\left(J_{n}+iK_{n}\right)-A_{n}\cos\sigma_{n}\left(V_{0}-iU_{0}\right)\\
& & -iA_{n}\sin\sigma_{n}\left(I_{0}+Q_{0}\right)\,,\\
L_{n+1}+iM_{n+1} & \simeq & e^{-2i\Delta}\left(L_{n}+iM_{n}\right)-A_{n}\sin\sigma_{n}\left(V_{0}+iU_{0}\right)\\
& & +iA_{n}\cos\sigma_{n}\left(I_{0}-Q_{0}\right)\,,\end{eqnarray*}
where we have assumed there is no initial chameleon flux which necessarily
sets $\vert\chi\rangle\sim\mathcal{O}(A)$. Note that $A$ depends on the magnitude of the transverse component of the magnetic
field, which fluctuates over the different domains, hence the subscript
$n$, whereas $\Delta$ is constant along the path save for
a weak dependence on the scale factor. However, since $\vert\Delta\vert\gg1$,
this will be averaged out across the path length. Solving this system
of equations for propagation through $N$ domains, we find the modified
Stokes parameters: \begin{eqnarray*}
I_{\gamma} & \simeq & \textstyle \left(1-\frac{1}{2}\mathcal{P}_{N}-\vartheta_{N}^{c-}\right)I_{0}+\left(\frac{1}{2}\kappa_{N}^{c}+\vartheta_{N}^{c+}\right)Q_{0}\\
& & +\left(\frac{1}{2}\kappa_{N}^{s}+\vartheta_{N}^{s+}\right)U_{0}-\varrho_{N}^{s-}V_{0}\,,\\
Q & \simeq & \textstyle \left(1-\frac{1}{2}\mathcal{P}_{N}-\Delta^{2}\left[\frac{1}{2}\eta_{N}^{ss}+\mu_{N}^{ss}\right]-\vartheta_{N}^{c-}\right)Q_{0}\\
& & +\left(\Delta^{2}\left[\frac{1}{2}\eta_{N}^{sc}+\mu_{N}^{cs}\right]-\vartheta_{N}^{s-}\right)U_{0}\\
 &  & +\left(\frac{1}{2}\kappa_{N}^{c}+\vartheta_{N}^{c+}\right)I_{0}+\left(\kappa_{N}^{s}\Delta+\varrho_{N}^{s+}\right)V_{0}\,,\\
U & \simeq & \textstyle \left(1-\frac{1}{2}\mathcal{P}_{N}-\Delta^{2}\left[\frac{1}{2}\eta_{N}^{cc}+\mu_{N}^{cc}\right]-\vartheta_{N}^{c-}\right)U_{0}\\
& & +\left(\Delta^{2}\left[\frac{1}{2}\eta_{N}^{sc}+\mu_{N}^{sc}\right]+\vartheta_{N}^{s-}\right)Q_{0}\label{eq:Ulinear}\\
 &  & +\left(\frac{1}{2}\kappa_{N}^{s}+\vartheta_{N}^{s+}\right)I_{0}-\left(\kappa_{N}^{c}\Delta+\varrho_{N}^{c+}\right)V_{0}\,,\\
V & \simeq & \textstyle \left(1-\frac{1}{2}\mathcal{P}_{N}-\Delta^{2}\left[\frac{1}{2}\mathcal{Q}_{N}+\mu_{N}^{cc}-\mu_{N}^{ss}\right]-\vartheta_{N}^{c-}\right)V_{0}\\
 &  & -\left(\kappa_{N}^{s}\Delta+\varrho_{N}^{s+}\right)Q_{0}+\left(\kappa_{N}^{c}\Delta+\varrho_{N}^{c+}\right)U_{0}-\varrho_{N}^{s-}I_{0}\,, \end{eqnarray*}
where we have defined
\begin{eqnarray*}
\mathcal{P}_{N}\equiv\textstyle\sum_{n=0}^{N-1}A_{n}^{2}\,, &  & \mathcal{Q}_{N}\equiv\textstyle\sum_{n=0}^{N-1}A_{n}^{4}\,,\end{eqnarray*}
and \begin{eqnarray*}
\vartheta_{N}^{c\pm} & = & \textstyle\sum_{n=0}^{N-1}\sum_{r=0}^{n-1}A_{n}A_{r}\cos\left(2\Delta(n-r)\right)\cos\left(\sigma_{r}\pm\sigma_{n}\right)\,,\\
\vartheta_{N}^{s\pm} & = &  \textstyle\sum_{n=0}^{N-1}\sum_{r=0}^{n-1}A_{n}A_{r}\cos\left(2\Delta(n-r)\right)\sin\left(\sigma_{r}\pm\sigma_{n}\right)\,,\\
\varrho_{N}^{c\pm} & = &  \textstyle\sum_{n=0}^{N-1}\sum_{r=0}^{n-1}A_{n}A_{r}\sin\left(2\Delta(n-r)\right)\cos\left(\sigma_{r}\pm\sigma_{n}\right)\,,\\
\varrho_{N}^{s\pm} & = & \textstyle \sum_{n=0}^{N-1}\sum_{r=0}^{n-1}A_{n}A_{r}\sin\left(2\Delta(n-r)\right)\sin\left(\sigma_{r}\pm\sigma_{n}\right)\,,\end{eqnarray*}
and \begin{eqnarray*}
\mu_{N}^{cc}& = & \textstyle\sum_{n=0}^{N-1}\sum_{r=0}^{n-1}A_{n}^{2}A_{r}^{2}\cos2\sigma_{r}\cos2\sigma_{n}\,,\\
\mu_{N}^{sc}& = & \textstyle\sum_{n=0}^{N-1}\sum_{r=0}^{n-1}A_{n}^{2}A_{r}^{2}\sin2\sigma_{r}\cos2\sigma_{n}\,,\\
\mu_{N}^{cs} & = & \textstyle\sum_{n=0}^{N-1}\sum_{r=0}^{n-1}A_{n}^{2}A_{r}^{2}\cos2\sigma_{r}\sin2\sigma_{n}\,,\\
\mu_{N}^{ss}& = & \textstyle\sum_{n=0}^{N-1}\sum_{r=0}^{n-1}A_{n}^{2}A_{r}^{2}\sin2\sigma_{r}\sin2\sigma_{n}\,,\end{eqnarray*}
and 
\begin{eqnarray*}
\kappa_{N}^{c}=\textstyle\sum_{n=0}^{N-1}A_{n}^{2}\cos2\sigma_{n}\,, &  & \kappa_{N}^{s}= \textstyle\sum_{n=0}^{N-1}A_{n}^{2}\sin2\sigma_{n}\,,\\
\eta_{N}^{ss}=\textstyle\sum_{n=0}^{N-1}A_{n}^{4}\sin^{2}2\sigma_{n}\,, &  & \eta_{N}^{cc}= \textstyle\sum_{n=0}^{N-1}A_{n}^{4}\cos^{2}2\sigma_{n}\,,\\
\eta_{N}^{sc}& = & \textstyle\sum_{n=0}^{N-1}A_{n}^{4}\sin2\sigma_{n}\cos2\sigma_{n}\,. \end{eqnarray*}
We define $z_{n}$ to be the location of the $n^{\mathrm{th}}$ domain, then $B_{x}(z_{n})=B_{n}\cos\sigma_{n}$ and $B_{y}(z_{n})=B_{n}\sin\sigma_{n}$ by definition. Remembering $A_{n}\equiv\omega_{0}B_{n}/p_{0}M_{\mathrm{eff}}$, we find
that the average over many lines of sight of the above quantities
are\begin{eqnarray*}
\textstyle\left\langle \mathcal{P}_{N}\right\rangle &= & \textstyle N\left(\frac{2\omega_{0}}{p_{0}M_{\mathrm{eff}}}\right)^{2}R_{\mathrm{B}}(0)\,,\\
\textstyle\left\langle \mathcal{Q}_{N}\right\rangle & = & \textstyle 4N\left(\frac{2\omega_{0}}{p_{0}M_{\mathrm{eff}}}\right)^{4}\left[R_{\mathrm{B}}(0)\right]^{2}\,,\end{eqnarray*}
\begin{eqnarray*}
\left\langle \vartheta_{N}^{c-}\right\rangle  & = &\textstyle \left(\frac{2\omega_{0}}{p_{0}M_{\mathrm{eff}}}\right)^{2}\displaystyle \sum_{n=0}^{N-1}\sum_{r=0}^{n-1}\cos2\Delta(n-r)R_{\mathrm{B}}(z_{r}-z_{n})\,,\\
\left\langle \varrho_{N}^{c-}\right\rangle  & = & \textstyle \left(\frac{2\omega_{0}}{p_{0}M_{\mathrm{eff}}}\right)^{2}\displaystyle \sum_{n=0}^{N-1}\sum_{r=0}^{n-1}\sin2\Delta(n-r)R_{\mathrm{B}}(z_{r}-z_{n})\,,\\
\left\langle \mu_{N}^{ss}\right\rangle & = & \left\langle \mu_{N}^{cc}\right\rangle = \textstyle \left(\frac{2\omega_{0}}{p_{0}M_{\mathrm{eff}}}\right)^{4}\displaystyle\sum_{n=0}^{N-1}\sum_{r=0}^{n-1}\left[R_{\mathrm{B}}(z_{r}-z_{n})\right]^{2}\,,\\
\left\langle \eta_{N}^{ss}\right\rangle & = & \left\langle \eta_{N}^{cc}\right\rangle =\textstyle N\left(\frac{2\omega_{0}}{p_{0}M_{\mathrm{eff}}}\right)^{4}\left[R_{\mathrm{B}}(0)\right]^{2}\,,\end{eqnarray*}
and all other quantities average to zero. We have assumed fluctuations in the magnetic field are approximately Gaussian so that the four-point correlations can be expressed in terms of the two-point correlation function, $R_{\mathrm{B}}(x-y)$, defined in Eq. (\ref{RB}).

Taking the average of the modified Stokes parameters, over the whole sky, we find

\begin{eqnarray*}
\frac{\langle\delta I_{\gamma}\rangle}{\langle I_{0}\rangle} & \simeq & -\left(\frac{2\omega_{0}}{p_{0}M_{\mathrm{eff}}}\right)^{2}\left(\frac{1}{2}NR_{\mathrm{B}}(0)\right.\\
& & \left.+\sum_{n=0}^{N-1}\sum_{r=0}^{n-1}\cos2\Delta(n-r)R_{\mathrm{B}}(z_{r}-z_{n})\right)\,,
\end{eqnarray*}
\begin{eqnarray*}
\frac{\langle\delta Q\rangle}{\langle Q_0\rangle} & \simeq & \frac{\langle\delta U\rangle}{\langle U_0\rangle}\\ 
&\simeq & -\left(\frac{2\omega_{0}}{p_{0}M_{\mathrm{eff}}}\right)^{2}\left(\frac{1}{2}NR_{\mathrm{B}}(0)\right.\\
& & \left.+\sum_{n=0}^{N-1}\sum_{r=0}^{n-1}\cos2\Delta(n-r)R_{\mathrm{B}}(z_{r}-z_{n})\right)\\
& & -\Delta^{2}\left(\frac{2\omega_{0}}{p_{0}M_{\mathrm{eff}}}\right)^{4}\left(\frac{1}{2}N\left[R_{\mathrm{B}}(0)\right]^{2}\right.\\
& & \left.+\sum_{n=0}^{N-1}\sum_{r=0}^{n-1}\left[R_{\mathrm{B}}(z_{r}-z_{n})\right]^{2}\right)\,,
\end{eqnarray*}
\begin{eqnarray*}
\frac{\langle\delta V\rangle}{\langle V_0\rangle} & \simeq & -\left(\frac{2\omega_{0}}{p_{0}M_{\mathrm{eff}}}\right)^{2}\left(\frac{1}{2}NR_{\mathrm{B}}(0)\right.\\
& & \left.+\sum_{n=0}^{N-1}\sum_{r=0}^{n-1}\cos2\Delta(n-r)R_{\mathrm{B}}(z_{r}-z_{n})\right) \\
& & -2\Delta^{2}\left(\frac{2\omega_{0}}{p_{0}M_{\mathrm{eff}}}\right)^{4}N\left[R_{\mathrm{B}}(0)\right]^{2}.\end{eqnarray*}
These results are quoted in section \ref{multiple_evolution}.


\begin{thebibliography}{10}
\bibitem{Kronberg94}P.P. Kronberg, Rept. Prog. Phys. $\mathbf{57}$, 325-382 (1994); C. Carilli and G. Taylor, Ann. Rev. Astron. Astrophys. $\mathbf{40}$, 319 (2001); L.M. Widrow, Rev. Mod. Phys. $\mathbf{74}$, 775-823 (2002); R.M. Kulsrud and E.G. Zweibel, Rept. Prog. Phys. $\mathbf{71}$, 0046091 (2008).
\bibitem{Ratra92}M.S. Turner and L.M Widrow, Phys. Rev. D{\bf 37}, 2743 (1988); B. Ratra, Astrophys. J. $\mathbf{391}$, L1-L4 (1992); M. Gasperini, M. Giovannini and G. Veneziano, Phys. Rev. D{\bf 52}, 6651-6655 (1995); O. Bertolami and D.F. Mota, Phys. Lett. B{\bf 455}, 96-103 (1999).
\bibitem{Kunze10}K.E. Kunze, Phys. Rev. D{\bf 81}, 043526 (2010); T. Kahniashvili $et\:al.$, $\mathtt{1004.3084[astro-ph]}$.
\bibitem{Tashiro06}H. Tashiro, N. Sugiyama and R. Banerjee, Phys. Rev. D$\mathbf{73}$, 023002 (2006); T. Kahniashvili, A.G. Tevzadze and B. Ratra, $\mathtt{0907.0197[astro-ph]}$.
\bibitem{Kosowsky09}T. Kahniashvili, Y. Maravin and A. Kosowsky, Phys. Rev. D$\mathbf{80}$, 023009 (2009).
\bibitem{Yamazaki10}D.G. Yamazaki, K. Ichiki, T. Kajino and G.J. Mathews, Phys. Rev. D$\mathbf{81}$, 023008 (2010).
\bibitem{Paoletti10}D. Paoletti and F. Finelli, $\mathtt{1005.0148[astro-ph]}$.
\bibitem{Khoury04}J. Khoury and A. Weltman, Phys. Rev. Lett. $\mathbf{93}$, 171104 (2004); Phys. Rev. D$\mathbf{69}$, 044026 (2004).
\bibitem{Brax07}P. Brax, C. van de Bruck, A.C. Davis, Phys. Rev. Lett. {\bf 99}, 121103, (2007); P. Brax, C. van de Bruck, A.C. Davis, D.F. Mota and
D. Shaw, Phys. Rev. D$\mathbf{76}$, 085010 (2007); C. Burrage, Phys. Rev. D$\mathbf{77}$, 043009 (2008).
\bibitem{Burrage08}C. Burrage, A.C. Davis and D.J. Shaw, Phys. Rev. D$\mathbf{79}$, 044028 (2009).
\bibitem{Brax09}P. Brax, C. van de Bruck, A.C. Davis and D. Shaw,	$\mathtt{0911.1086[astro-ph]}$; G.G. Raffelt, Lect. Notes Phys. {\bf 741}, 51 (2008); D.F. Mota and D.J. Shaw, Phys. Rev. Lett. {\bf 97} 151102 (2006); D.F. Mota and D.J. Shaw, Phys. Rev. D.{\bf 75}, 063501  (2007).
\bibitem{Davis09}A.C. Davis, C.A.O. Schelpe and D.J. Shaw, Phys. Rev. D$\mathbf{80}$, 064016 (2009).
\bibitem{Olive08}K.A. Olive and M. Pospelov, Phys. Rev. D{\bf 77}, 043524 (2008); D.F. Mota and J.D. Barrow, Phys. Lett. B{\bf 581}, 141-146 (2004); D.F. Mota and J.D. Barrow, Mon. Not. Roy. Astron. Soc. $\mathbf{349}$, 291 (2004).
\bibitem{Mirizzi09}A. Mirizzi, J. Redondo and G. Sigl, JCAP $\mathbf{0908}$, 001 (2009).
\bibitem{Barrow98} K. Jedamzik, V. Katalinic and A.V. Olinto, Phys. Rev. D$\mathbf{57}$, 3264-3284 (1998); K. Subramanian and J.D. Barrow, Phys. Rev. Lett. $\mathbf{81}$, 3575-3578 (1998).
\bibitem{Agarwal09}N. Agarwal, A. Kamal and P. Jain, $\mathtt{0911.0429[hep-ph]}$.    
\bibitem{Barrow07}J.D. Barrow, R. Maartens and C.G. Tsagas, Phys. Rept. $\mathbf{449}$, 131-171 (2007). 
\bibitem{Raffelt88}G. Raffelt and L. Stodolsky, Phys. Rev. D {\bf 37} 1237 (1988). 
\bibitem{Seager99}S. Seager, D.D. Sasselov and D. Scott, Astrophys. J. Suppl. $\mathbf{128}$, 407-430 (2000); A. Lewis, J. Weller and R. Battye, Mon. Not. Roy. Astron. Soc. $\mathbf{373}$, 561-570 (2006).
\bibitem{Steigman06}G. Steigman, JCAP $\mathbf{0610}$, 016 (2006).
\bibitem{Dunkley08}WMAP Collaboration, G. Hinshaw $et\:al.$, Astrophys. J. Suppl. $\mathbf{180}$, 225 (2009).
\bibitem{Kosowsky05}A. Kosowsky, T. Kahniashvili, G. Lavrelashvili and B. Ratra, Phys. Rev. D$\mathbf{71}$, 043006 (2005); A. Mack, T. Kahniashvili and A. Kosowsky, Phys. Rev. D$\mathbf{65}$, 123004 (2002).
\bibitem{Caprini01}C. Caprini and R. Durrer, Phys. Rev. D$\mathbf{65}$, 023517 (2001).
\bibitem{Schelpe10}C.A.O. Schelpe, in preparation.
\bibitem{Fixsen96}D.J. Fixsen $et\:al.$, Astrophys. J. $\mathbf{473}$, 576 (1996).
\bibitem{Fixsen02}D. J. Fixsen and J.C. Mather, Astrophys. J. $\mathbf{581}$, 817-822 (2002); J.C. Mather $et\:al.$, Astrophys. J. $\mathbf{512}$, 511-520 (1999).
\end{thebibliography}
\end{document}